\documentclass[pra,twocolumn,notitlepage]{revtex4-1}
\usepackage{amsmath}
\usepackage{amsfonts}
\usepackage{amssymb}
\usepackage{graphicx}
\usepackage{color}

\begin{document}

\title{The reference system and not completely positive open quantum dynamics}

\author{Linta Joseph}
\email{linta.joseph.gr@dartmouth.edu}
\affiliation{Department of Physics and Astronomy, Dartmouth College, 6127 Wilder Laboratory, Hanover, New Hampshire 03755, USA}

\author{Anil Shaji}
\email{shaji@iisertvm.ac.in}
\affiliation{School of Physics, Indian Institute of Science Education and Research Thiruvananthapuram, India 695016}

\begin{abstract}
Open quantum dynamics in a tripartite scenario including a system, its environment and a passive reference is shown to resolve several open questions regarding not completely positive (NCP) dynamical maps as valid descriptions of open quantum evolution. The steering states of the system and the environment with respect to the reference, reduced down to a dense, compact set of states of the system alone, provides a well defined domain of action for a bonafide dynamical map describing the open evolution of the system. The map is not restricted to being completely positive (CP) but it preserves the positivity of all states in its domain. NCP open dynamics corresponding to different initial configurations of the tripartite system are explored. 
\end{abstract}


\maketitle

\section{Introduction}

Information delocalized across multiple, identifiable, subsystems is a feature of quantum mechanics that leads to several subtle effects in the open dynamics of the individual subsystems. It seems increasingly clear that the flow of information may be just as important as the flow of matter, energy and momentum in determining the dynamics of a quantum system evolving in contact with a quantum environment~\cite{breuer02,Lloyd:2013wi,Ingarden:vw}. Such dynamics with low dimensional quantum systems evolving in contact with a micro or mesoscopic environment having quantum features can be a theoretical as well as experimental test bed for the interplay between information flows and mechanics. In this Paper we consider the connection between information flows and the nature as well as domain of action of dynamical maps that describe finite time open evolution of a quantum system. We see that information that may lie delocalized across the system, its environment and the rest of the universe prior to the start of the particular open dynamics of the system that is of interest can have a bearing on the nature of the observed dynamics. 

While the time evolved state of an isolated quantum system is given by an unitary transformation acting on its initial state, the corresponding transformation for an open quantum system is described by a dynamical map~\cite{ecgs61b} defined from the set of density matrices to itself. The map has to be trace preserving, hermiticity preserving and its action should be positivity preserving on the set of states it is defined on. However, a stronger condition of complete positivity~\cite{choi72} is often proposed as being required of the map~\cite{kraus71a,kraus83,nielsen00,stinespring55}. The most widely accepted argument for complete positivity of the map involves the introduction of an arbitrary `blind' and `dead' reference or witness system~\cite{schumacher96a} which does not interact with the system of interest when it is evolving in contact with its environment. A dynamical map acting on the system can potentially transform the density matrices corresponding to certain joint states of the system and the reference into matrices that are not positive if the map is not CP. This potential pitfall is taken as one of the reasons to assert that reduced dynamics of the system should be described in terms of CP maps exclusively. If the initial state of the system $S$ and its environment $E$ is a product state of the form $\rho_{S} \otimes \eta_{E}$, where $\eta_E$ is a fixed state of $E$, the reduced dynamics induced by the joint unitary evolution of $S$ and $E$ is CP. In the presence of initial system-environment correlations, the reduced dynamics is however not necessarily CP~\cite{jordan04,jordan04b,pechukas94,pechukas95,shaji05a}. Is complete positivity really required and, if so, what kind of initial correlations between the system and environment, in general, guarantee CP reduced dynamics? A complete answer to this question is still not forthcoming~\cite{Brodutch:2013bx,Dominy:2015fw,rodriguez_completely_2008,Shabani:2009dy,Shabani:2016if}.

By considering the role of the reference system $R$ in greater detail and by using an information theoretic framework involving a tripartite approach in place of the usual bipartite one allowed Buscemi~\cite{Buscemi:2014dc} to identify a more general set of conditions under which the reduced dynamics is CP as well as to define the domain of action of such a CP dynamical map. In~\cite{Buscemi:2014dc}, the observation that NCP reduced dynamics may lead to violation of (classical) data and energy processing principles~\cite{Breuer:2009cu,DArrigo:2014jd,Jennings:2010kr,Partovi:2008if,Xu:2013ia,Rivas:2014bl} is leveraged to recover CP dynamics by assuming that the quantum data processing inequality (DPI) always holds. When the initial tripartite state of $R$, $S$ and $E$, denoted as $\rho_{\!R\!S\!E}$, constitutes a short Markov chain with $I(R:E|S) = 0$ the reduced dynamics on $S$ is always CP and the dynamics does not violate the quantum DPI. Here $I(R:E|S)$ is the quantum mutual information between $R$ and $E$ conditioned on $S$ with
 \[ I(A:B) \equiv S(\rho_{A}) + S(\rho_{B}) - S(\rho_{\!A\!B}),\] 
 and $S(\rho) = -{\rm tr}(\rho \log \rho)$. The quantum mutual information quantifies all the correlations including the delocalised quantum information shared across the two systems.  When  $I(R:E|S) = 0$ the reduced dynamics on $S$ induced by arbitrary unitary dynamics of the $S\!E$ subsystem is always CP and the dynamics does not violate the quantum DPI. The quantum DPI is the condition, 
 \[ I(R:S) \geq I(R:S'), \] 
 where $I(R:S)$ is the quantum mutual information between $R$ and $S$ before the open evolution of $S$ in contact with $E$ and $I(R:S')$ is the mutual information after. The domain of action of the CP dynamical map is the reduced steering set of $S$ obtained by tracing out $E$ from the set of steering states of $S\!E$ due to $R$~\cite{Wiseman:2007ca,Cavalcanti:2017ba}. 
 
 The assumptions that the initial tripartite state forms a short Markov chain and that the quantum DPI is satisfied by the dynamics are not physically mandated or motivated. Relaxing them puts into context several of the recent works attempting to understand the role of initial correlations between $S$ and $E$ in open system dynamics~\cite{jordan04,jordan04b,pechukas94,pechukas95,shaji05a,Brodutch:2013bx,Dominy:2015fw,rodriguez_completely_2008,Shabani:2009dy,Shabani:2016if,Modi:2012iv}. The key observation is that the reduced set of steering states of $S$ projected down from the tripartite $R\!S\!E$ system provides the natural set on which the open dynamics is to be considered when it is NCP. By bringing in the idea of quantum steering ellipsoids~\cite{Jevtic:2014fm} we are able to go significantly beyond the limited results in~\cite{Buscemi:2014dc} that focuses only on the conventional picture of CP dynamics. By going beyond the traditional confines of CP reduced dynamics we are able to see the role of delocalized quantum information shared between the three systems not only in producing NCP reduced dynamics but also in giving a natural explanation and domain of action for it. 
 
 The remainder of this Paper is structured as follows: In the next section we give a general proof of our main result. In section \ref{sec:discuss} we discuss our result putting it in context with respect to previous attempts to understand NCP open dynamics. Examples are numerically investigated in the subsequent section and our main findings are summarized in Sec.~\ref{sec:summary}.
 
\section{Domain of definition of NCP reduced dynamics \label{sec:main}}

The set of system-environment states that can be steered from a given joint state $\rho_{\!R\!S\!E}$ is given by:
\begin{equation}\label{eq:Steering}
\mathcal{S}_{SE}(\rho_{\!R\!S\!E}):=\left \{ \frac {\rm Tr_{R}[(\mathcal{P}_{R}\otimes \textbf{1}_{S}\otimes \textbf{1}_{E})\rho_{\!R\!S\!E}]} {{\rm Tr}[(\mathcal{P}_{R}\otimes \textbf{1}_{S}\otimes \textbf{1}_{E})\rho_{\!R\!S\!E}]}\right\} 
\end{equation}
where $\mathcal{P}_{R}\in\textbf{L}^{+}(\mathcal{H}_{R})$ is the set of all positive semi definite linear operators acting on $\mathcal{H}_{R}$, the Hilbert space of $R$. Let $R$, $S$ and $E$ be quantum systems with finite dimensional Hilbert spaces of dimensions $N_{R}$, $N_{S}$ and $N_{E}$ respectively. The $N_{J}^{2}-1$ generators of the special unitary group in $N_{J}$ dimensions, SU($N_{J}$), along with the identity matrix in $N_{J}$ dimensions form an operator basis in terms of which any complex $N_{J}\times N_{J}$ matrix can be expanded. We label the traceless generators of SU($N_{J}$) as $F_{J}^{i}$ where $J=R, S$ or $E$. The generators satisfy the commutation relations, 
\[ [F_{J}^{a}, \, F_{J}^{b}] = i f_{J}^{abc}F_{J}^{c}, \qquad a,b,c=1,2, \ldots, N_{J}^{2} - 1,\]
where $f_{J}^{abc}$ are the structure constants of SU($N_{J}$) which, in turn, are completely antisymmetric with respect to exchange of indices. We choose the normalisation of the generators so that
\[ (F_{J}^{a})^{2}= \openone_{J}, \]
where $\openone_{J}$ is the $N_{J} \times N_{J}$ unit matrix. In terms of the generators of SU($N_{J}$) we can write $\rho_{R\!S\!E}$ as:
\begin{eqnarray}
	\label{eq:rhoRSE2a}
	\rho_{\!R\!S\!E} & = & \frac{1}{N_{R}N_{S}N_{E}} \Big(  \openone_{R} \otimes \openone_{S} \otimes \openone_{E} \nonumber \\
	&& \quad  + a_{i} F_{R}^{i} \otimes \openone_{S} \otimes \openone_{E}  + e_{j} \openone_{R} \otimes F_{S}^{j} \otimes \openone_{E} \nonumber \\
	&& \quad + e_{n_{S}+k} \openone_{R} \otimes \openone_{S} \otimes F_{E}^{k} \nonumber \\
	&& \quad +e_{n_{S}+n_{E}+n_{S}(j-1)+k}\openone_{R} \otimes F_{S}^{j} \otimes F_{E}^{k} \nonumber \\
	&& \qquad   + T_{j,i} F_{R}^{i} \otimes F_{S}^{j} \otimes \openone_{E}  + T_{n_{S}+k,i} F_{R}^{i} \otimes \openone_{S} \otimes F_{E}^{k} \nonumber \\
	&& \qquad + T_{n_{S}+n_{E}+n_{S}(j-1)+k,i} F_{R}^{i} \otimes F_{S}^{j} \otimes F_{E}^{k}  \Big),
\end{eqnarray}
where $n_{J} \equiv N_{J}^{2} - 1$. Here $i = 1, \ldots, n_{R}$, $j=1,\ldots, n_{S}$ and $k=1,\ldots,n_{E}$. All the parameters defining the state can be packaged into a matrix as 
\begin{equation}
	\label{eq:theta1}
	\Theta = \left( \begin{array}{cc} 1 & \vec{a}^{T} \\ \vec{e} & T \end{array} \right),
\end{equation}
where $\Theta$ has $(n_{S}+1)(n_{E}+1)$ rows and $n_{R}+1$ columns. We can write an arbitrary positive operator on $R$ as 
\[ \hat{E} = X_{\mu} F_{R}^{\mu},\] 
with $X_{0}^{2} \geq \sum_{i} X_{i}^{2}$, $X_{0} > 0$ and the greek index $\mu$ taking on values $ 0, 1, \ldots, n_{R}$ with $F_{R}^{0} \equiv \openone_{R}$. Using the fact that $F_{R}^{i}$ are traceless matrices squaring to the identity operator, we see that the steering state of $S\!E$ corresponding to the application of $\hat{E}$ on $R$ is then given by the vector $\vec{e}^{X} = \Theta X$ as
\begin{eqnarray}
	\label{eq:rhoSEa2}
	\rho_{\!S\!E}^{X}&  = &  {\rm Tr}_{R} (\hat{E} \rho_{R\!S\!E})  \nonumber \\
	&= & \frac{1}{N_{S}N_{E}} \Big( e^{X}_{0} \openone_{S} \otimes \openone_{E} + e^{X}_{j} F_{S}^{j} \otimes \openone_{E}  \nonumber \\
	&& \qquad + e^{X}_{n_{S}+k} \openone_{S} \otimes F_{E}^{k} \nonumber \\
	&& \qquad + e^{X}_{n_{S}+n_{E}+n_{S}(j-1)+k} F_{S}^{j} \otimes F_{E}^{k}\Big) . 
\end{eqnarray} 
	with
\begin{eqnarray}
	\label{eq:appA12}
	e_{j}^{X} & = & e_{j} + T_{j,i}X_{i} = \Theta_{j, \mu}X_{\mu}, \nonumber \\
	e_{n_{S}+k}^{X} & = & e_{n_{S}+k} + T_{n_{S}+k,i}X_{i} = \Theta_{n_{S}+k,\mu}X_{\mu}, \nonumber \\
	e_{n_{S}+n_{E}+n_{S}(j-1)+k}^{X} & = & e_{n_{S}+n_{E}+n_{S}(j-1)+k} \nonumber \\
	&& \quad  + T_{n_{S}+n_{E}+n_{S}(j-1)+k,i}X_{i} \nonumber \\
	& = &  \Theta_{n_{S}+n_{E}+n_{S}(j-1)+k,\mu}X_{\mu}.
\end{eqnarray}
Normalising the $S\!E$ reduced state and considering SLOCC (Stochastic local operations and classical communcations) transformations on $R$ that do not change the steering set on $SE$ lets us set $\vec{a} = 0$ and $X_{0}=1$ giving $e_{0}^{X} = 1$ as shown in~\cite{Jevtic:2014fm}. The reduced steering set of $S$ obtained by tracing out $E$ from $\rho_{SE}^{X}$ has states of the form
\begin{equation}
	\label{eq:rhoINI}
	 \rho_{S}^{X} = \frac{1}{N_{S}}(\openone_{S} + \Theta_{j, \mu}X_{\mu} F_{S}^{j}).
\end{equation}
The joint unitary evolution of $S$ and $E$ is represented in terms of its action on the operator basis furnished by the tensor products of SU($N_{J}$) generators (including $F_{S}^{0} \equiv \openone_{S}$ and $F_{E}^{0} \equiv \openone_{E}$) corresponding to each subsystem as 
\[ U F_{S}^{\zeta} \otimes F_{E}^{\eta} U^{\dagger} = u_{\alpha \beta}^{\zeta \eta} F_{S}^{\alpha} \otimes F_{E}^{\beta},\]
where $\zeta, \alpha, = 0,\ldots,n_{S}$, and  $\eta, \beta,  = 0,\ldots,n_{E}$. The unitarity condition $U^{\dagger}U  = UU^{\dagger} = \openone$ means that 
\[ u_{00}^{\alpha \beta}= u_{\alpha \beta}^{00} = \delta_{\alpha 0} \delta_{\beta 0}.\]
On application of the unitary, states in the steering set of $S\!E$ are transformed to
\begin{eqnarray*}
	\widetilde{\rho}_{SE}^{X}&  = & \frac{1}{4} \big( \openone_{S} \otimes \openone_{E} + e^{X}_{j} u^{j0}_{l \beta} F_{S}^{l} \otimes F_{E}^{\beta} + e^{X}_{j} u^{j0}_{0m} \openone_{S} \otimes F_{E}^{m} \nonumber \\
	&& \qquad + e^{X}_{n_{S}+k}  u^{0k}_{\alpha m} F_{S}^{\alpha} \otimes F_{E}^{m} + e^{X}_{n_{S}+k} u^{0k}_{l0} F_{S}^{l} \otimes \openone_{E}  \\
	 && \qquad + e^{X}_{n_{S}+n_{E}+n_{S}(j-1)+k} u^{jk}_{0 m} \openone_{S} \otimes F_{E}^{m} \nonumber \\
	 && \qquad + e^{X}_{n_{S}+n_{E}+n_{S}(j-1)+k} u^{jk}_{l 0} F_{S}^{l} \otimes \openone_{E}  \\
	 && \qquad  + e^{X}_{n_{S}+n_{E}+n_{S}(j-1)+k} u^{jk}_{l m} F_{S}^{l} \otimes F_{E}^{m}\big),
\end{eqnarray*} 
with $j,l=1,\dots, n_{S}$ and $k,m=1,\ldots,n_{E}$. Tracing out $E$ from the state above (keeping only the terms with $\openone_{E} = F_{E}^{0}$), we obtain the states in the reduced steering set of $S$ as 
\begin{eqnarray*}
	\widetilde{\rho}_{S}^{X} & = &  \frac{1}{2} \Big[ \openone_{S} + \big( e^{X}_{l} u^{l0}_{j 0} +e^{X}_{n_{S}+k} u^{0k}_{j0} \\
	&& \quad +e^{X}_{n_{S}+n_{E}+n_{S}(l-1)+k} u^{lk}_{j 0}  \big) F_{S}^{j} \Big], 
\end{eqnarray*}
after exchanging the summed over indices $j$ and $l$. Using Eq.~(\ref{eq:appA12}), we can write the coefficient of $F_{S}^{j}$ in the above equation as $\widetilde{\Theta}_{j,\mu} X_{\mu}$ where,
\[  \widetilde{\Theta}_{j,\mu} =u^{l0}_{j 0} \Theta_{l,\mu}+ u^{0k}_{j0} \Theta_{n_{S}+k, \mu}+ u^{lk}_{j 0} \Theta_{n_{S}+n_{E}+n_{S}(l-1)+k, \mu}. \]
So we have 
\begin{equation}
	 \rho_{S}^{X} \rightarrow \widetilde{\rho}_{S}^{X} = \frac{1}{2} \big[ \openone_{S} + \widetilde{\Theta}_{j,\mu} X_{\mu} F_{S}^{j} \big]. 
	 \label{eq:main1}
\end{equation}
This is the central result of this Paper. We can get an intuitive understanding of the Eq.~(\ref{eq:main1}) by noting that the projection operators on $R$ characterized by $X_\mu$ commute with $U$ acting as $\openone_{R} \otimes U_{S\!E}$ on $\rho_{\!R\!S\!E}$. So the transformation induced by $U$ on each point of the reduced steering set of $S$ is independent of the particular $X_\mu$  from which it came and the initial state of $S$ in Eq.~(\ref{eq:rhoINI}) is connected to the final state in Eq.~(\ref{eq:main1}) by the transformation
\begin{equation}
	\label{eq:ThetaTransform}
	\Theta_{j,\mu} \rightarrow \widetilde{\Theta}_{j,\mu}.
\end{equation} 
The matrix $\Theta$, as mentioned earlier, is a way of writing the joint state of $R$, $S$ and $E$. It follows that Eq.~(\ref{eq:ThetaTransform}) is essentially the transformation of the tripartite $R\!S\!E$ state due to the action of $\openone_{R} \otimes U_{S\!E}$. The action of the induced dynamical map on $S$ is then to transform each point in the reduced steering set generated from the tripartite state $\Theta$ to the corresponding point characterized by the same $X_\mu$ in the reduced steering set of $\widetilde{\Theta}$. Both reduced steering sets are subsets of the set of all states of $S$ and significantly the one corresponding to the initial tripartite state $\rho_{R\!S\!E}$ is the domain on which the dynamical map induced by unitary $S\!E$ evolution is defined irrespective of whether the map is CP or NCP. 

\section{The role of the reference system \label{sec:discuss}}

The main result contained in Eq.~(\ref{eq:main1}) is significant because it answers a question that has lingered ever since NCP open dynamics was considered as a possibility backed by experimental evidence~\cite{pechukas94}. If the condition that the system and the environment is initially in a product state of the form $\rho_{S} \otimes \rho_{E}$, which is a necessary and sufficient condition for complete positivity of the reduced dynamics of $S$, is relaxed, then is there even a valid dynamical map that can be defined consistently? For the description of open quantum dynamics in terms of a map to be useful in any sense, it has to apply to a compact and dense set of states of $S$. It should be possible to pick a sufficient number of linearly independent states from this set to do quantum process tomography~\cite{nielsen00,Chuang:2009ie,YuenZhou:2014js} and reconstruct the dynamical map. In the presence of initial correlations between $S$ and $E$ it was not clear whether such a set of states exists. Even if it did exist how would one define it?

In~\cite{jordan04}, it was shown that if the initial state of $S\!E$ is an entangled two qubit state, then the transformations on the states of $S$ induced by unitary evolution of $S\!E$ is NCP in general. {\textcolor{black} {However, starting from a particular initial state of $S\!E$ we obtain only the transformation of a particular state of $S$. Such a transformation does not constitute a map but only a ``point transformation''.}} Such point transformations obtained could be elevated to the status of a dynamical map acting on a well defined, dense subset of states of $S$ because it was positivity preserving on all states of $S$ that were `consistent' with the specification of the initial entanglement between $S$ and $E$ irrespective of the choice of unitary evolution of $S\!E$. In the present notation, this translates to the statement that once the coefficients $e^{X}_{n_S+n_E+n_S(j-1)+k}$ are fixed then the coefficients $e^{X}_{j}$ that define the reduced state of $S$ cannot take all the possible values such that $\sum_{j} (e_{j}^{X})^{2} \leq 1$, while still keeping $\rho^{X}_{SE}$ positive. It was precisely on this subset of states corresponding to allowed values of $e_{j}^{X}$ - termed the `compatibility domain' - that the transformation was positivity preserving and it could be elevated to a dynamical map defined on this domain. However, several open questions remained. For instance, it is possible to choose $e^{X}_{n_S+n_E+n_S(j-1)+k}$ such that $e^{X}_{j}$ would also be uniquely fixed. In this case only a point transformation acting on a single system state can be obtained and not a map. 

Another serious conceptual issue that remained was that for every choice of $e^{X}_{j}$ consistent with some fixed $e^{X}_{n_S+n_E+n_S(j-1)+k}$, the coefficients determining the state of $E$, $e^{X}_{n_E+k}$ can also end up being automatically constrained. This would mean that for each state of $S$ in the domain of action of the map, the corresponding initial state of the environment would have to be chosen in a manner that depends on the state of $S$ in order to elevate a point transformation of interest to a map. There was no clear justification for assuming that states of the environment arise in a manner that depend on the system state just so that an NCP dynamical map with a well defined domain of action would be observed in an experiment. 

Approaching the problem using steering states provides a clear, elegant and aesthetically pleasing resolution to the question whether there can be a consistent mathematical definition of NCP dynamics with an unambiguous physical interpretation. An arbitrary unitary acting on any state in the steering set of $S\!E$ will induce a transformation that is positivity preserving on the corresponding reduced steering set of $S$ obtained by tracing out $E$. Furthermore, a given unitary acting on $S\!E$ steering set will induce the same transformation on all the states in the reduced set of $S$ as can be seen from Eq.~(\ref{eq:ThetaTransform}). The set of states on which the transformation acts is dense and compact which means that the transformation can be treated as a bonafide dynamical map with a well defined domain. Operationally this also means that one can reconstruct the map from observing the transformation occurring to  a sufficient, finite number of linearly independent initial states of $S$ and using the techniques of quantum process tomography~\cite{nielsen00,Chuang:2009ie}. Since the steering set of $S\!E$ from which the domain of action of the maps follows is itself derived from a {\em single} state of $R\!S\!E$, there is no longer any mystery in having the initial state of the environment dependent on the state of $S$. The choice of projection ($X_\mu$) on $R$ that does the steering determines the states of both $S$ and $E$ together and associated with each state of $S$ there can be unique states of $E$.  Where $R$ remains independent of both $S$ and $E$ in $\rho_{\!R\!S\!E}$, no steering is possible and so no dynamical map is obtained. In a quantum tomography setting such a scenario cannot arise since, at the very least, the role of $R$ is taken by the instrument or agency that initializes the states of $S$ as discussed below.  

It is useful to imagine that $R$ is a preparation device~\cite{Modi:2012iv} which, like any good 'reference' system, by definition, does not interact with $S$ during the open dynamics of interest when $S$ and $E$ are interacting. $R$ necessarily interacts with $S$ prior to its open evolution but with respect to its interaction with $E$ we have the following four cases:
\begin{enumerate}
	\item $R$ and $E$ do not interact with each other at all at any point in time.
	\item $R$ does not interact with $E$ prior to the preparation of the initial state of $S$ but subsequently while the joint evolution of $S$ and $E$ is happening, $R$ is  interacting with $E$.
	\item $R$ and $E$ interact prior to the preparation of the initial state of $S$ but not afterward. 
	\item $R$ and $E$ interact both before and after the  preparation of the initial state of $S$. 
\end{enumerate}

In all of the above four cases no assumption is made about the initial state of the $S\!E$ subsystem and the two could very well be in an entangled state even after $R$ has performed the state preparation on $S$. In the first two cases, $R$, $S$ and $E$ form short Markov chains. Initially there is no delocalized quantum information shared between $R$ and $E$. However in the second case, the subsequent dynamics is not guaranteed to satisfy the quantum DPI because the interaction between $R$ and $E$ during the joint $S\!E$ evolution can produce a flow of information from $R$ to $S$ through $E$ in those cases where $R$ remains entangled with $S$ even after preparation. So it appears as though even if the initial $S$, $E$ state is a factorized state one may be able to obtain NCP reduced dynamics for $S$. However this is a false positive since the joint effect of the $S\!E$ interaction during the dynamics of interest and the $R\!E$ interaction is to generate a coupling between $S$ and $R\!E$. So, in this case, even if $R$ and $E$ happen to be distinct physical systems, one of which that interacts with $S$ only during the initial state preparation and the other only after the initial state preparation, $R\!E$ has to be treated as the environment of $S$. Since $S$ can have shared nonClassical correlations with this composite environment at the point of time when the initial state preparation is deemed to be done, the subsequent dynamics can be NCP~\cite{jordan04}. The role of the true reference system will have to be played by the rest of the universe in this case. 

In case 3, the initial state of $R\!S\!E$ does not necessarily constitute a short Markov chain and so one expects the reduced dynamics of $S$ to be NCP. Indeed the joint effect of the $R\!S$ and $R\!E$ interactions before and during the preparation of the initial state of $S$ typically puts $S\!E$ in a state in which there are initial shared nonClassical correlations between the two subsystems. Case 4 is another false positive that needs to be discounted because $R$ again effectively becomes a part of the environment of $S$ during the open evolution and so we cannot treat it either as a valid preparation device or as a good reference system.  

Treating $R$ as the preparation devices lets us also see the kind of results one should expect in a quantum process tomography experiment. The initial $S\!E$ state is of the form $\rho_{S} \otimes \sigma_{E}$ only if $R$ and $S$  do not interact with $E$ during the preparation. This is possible only case 1, once the false positive given by case 2 wherein $R$ is both preparation device and environment is discounted. An alternate possibility is that $E$ can be a large `thermal' system on which interaction with $R$ and $S$ have negligible effect. Even if the initial state is $\rho_{S} \otimes \sigma_{E}$, the preparation device $R$ has to be perfect and ideal for it to initialize or {\em steer} $S$ into any one of the states in its Hilbert space. However with an imperfect preparation device, one may still get sufficient number of linearly independent initializations of $S$ to reconstruct a dynamical map which in turn will be CP. 

A restricted version of case 3 is considered in~\cite{Buscemi:2014dc} where the preparation device can produce entangled initial states of $S\!E$. However at the end of the preparation procedure, $R$ has no residual correlations with $E$ even if it has correlations with $S$ so that $R\!S\!E$ forms a short Markov chain. The quantum DPI is satisfied by the subsequent dynamics in case 3 because $R$ and $E$ do not interact after preparation. The reduced dynamics is CP in this case but its domain is limited to the reduced steering set of $S$. Case 3 is more general in the sense that $R\!S\!E$ need not form a short Markov chain at the time of preparation of the initial state of $S$. Our key result is that the NCP map or a map that is not positive even that will be obtained through process tomography involving such preparations still has a valid interpretation and a well defined domain of action. It may be noted that case 3 is considered through an alternative but relatively cumbersome approach in~\cite{Modi:2012iv} using the language of a preparation maps. 

When $S$ is a single quantum system and when we are concerned about its open dynamics over short time periods, its immediate environment is more often than not in itself microscopic or mesoscopic. The action of preparing an initial state of $S$ not affecting the state of $E$ in any way is an exceptional scenario in this case and in this sense so is CP reduced dynamics of the system. It is more reasonable to assume that right at the end of each preparation there is a certain amount of quantum information delocalized across $S$, $E$ and $R$, where $R$ is effectively the quantum parts of the preparing device and $I(R:E|S) \neq 0$. The device $R$, by definition, plays no further part in the joint evolution of $S$ and $E$. Furthermore, in any particular realization of the open dynamics, which, for instance, could be trials in a quantum process tomography experiment, $R$ is projected on to particular states by those processes that put preparing devices in human readable, classical states~\cite{Zurek:2003fm} and announce that a particular preparation has been done. This last step puts the $S\!E$ subsystem into one of its steering states. 

\section{Three qubit examples \label{sec:examples}}

Next we consider illustrative examples wherein $R$, $S$ and $E$ are qubits. The special case where $R$, $S$ and $E$ are qubits is obtained in a straightforward manner by replacing the generators $F_J^i$ with $\sigma_J^i$ where the $\sigma^i$'s are the Pauli spin operators and replacing the structure constants, $f^{abc}_J$ with the Levi-Civita symbol $\epsilon^{abc}$. The three qubit state $\rho_{\!R\!S\!E}$ can then be written in the form
\begin{eqnarray}
	\label{eq:rhoRSE2}
	\rho_{\!R\!S\!E} & = & \frac{1}{8} \big(  \openone_{R} \otimes \openone_{S} \otimes \openone_{E} + a_{i} \sigma_{R}^{i} \otimes \openone_{S} \otimes \openone_{E}  \nonumber \\
	&& \quad + e_{j} \openone_{R} \otimes \sigma_{S}^{j} \otimes \openone_{E} + e_{3+k} \openone_{R} \otimes \openone_{S} \otimes \sigma_{E}^{k} \nonumber \\
	&& \qquad +e_{6+3(j-1)+k}\openone_{R} \otimes \sigma_{S}^{j} \otimes \sigma_{E}^{k}   \nonumber \\
	&& \qquad + T_{j,i} \sigma_{R}^{i} \otimes \sigma_{S}^{j} \otimes \openone_{k}  + T_{3+k,i} \sigma_{R}^{i} \otimes \openone_{S} \otimes \sigma_{E}^{k} \nonumber \\
	&& \qquad + T_{6+3(j-1)+k,i} \sigma_{R}^{i} \otimes \sigma_{S}^{j} \otimes \sigma_{E}^{k}  \big)
\end{eqnarray}
with $i,j,k$ in this case running over the values $1,2,3$. The arbitrary positive operator on $R$ can be written as $\hat{E} = X_{\mu} \sigma_{R}^{\mu}$ with $X_0$ set to unity using the SLOCC freedom and $1= X_{0}^{2} \geq \sum_{i=1}^{3} X_{i}^{2}$. Here the greek indices runs over the four values $0,\ldots,3$ with $\sigma^0 \equiv \openone_{2 \times 2}$. The steering state of $S\!E$ corresponding to the application of $\hat{E}$ on $R$ is given by the $16 \times 1$ vector $\vec{e}^{X}=\Theta X$ as
\begin{eqnarray}
	\label{eq:rhoSE2}
	\rho_{\!S\!E}^{X}&  = & \frac{1}{4} \big( e^{X}_{0} \openone_{S} \otimes \openone_{E} + e^{X}_{j} \sigma_{S}^{j} \otimes \openone_{E} + e^{X}_{3+k} \openone_{S} \otimes \sigma_{E}^{k} \nonumber \\
	 && \qquad + e^{X}_{6+3(j-1)+k} \sigma_{S}^{j} \otimes \sigma_{E}^{k}\big). 
\end{eqnarray}
The reduced steering set of $S$ consists of the states
\begin{equation}
	\label{eq:steeringSi}
	\rho_{S}^{X} = \frac{1}{2} \big[ \openone_{S} + e_j^X\sigma_{S}^{j} \big], \quad e_j^X = \Theta_{j, \mu}X^\mu.
\end{equation}
Our main result in Eq.~(\ref{eq:main1}) tells us that joint evolution of the $S\!E$ system described by a unitary $U$ leads to the reduced system state,
\begin{equation}
	\label{eq:central1}
	\widetilde{\rho}_{S}^{X} = \frac{1}{2} \big[ \openone_{S} + \widetilde{\Theta}_{j,\mu} X_{\mu} \sigma_{S}^{j} \big],
\end{equation}
where
\begin{equation}
	\widetilde{\Theta}_{j,\mu} = u_{j0}^{l0} \Theta_{l,\mu} +u_{j0}^{0k} \Theta_{3+k,\mu} +u_{j0}^{lk} \Theta_{6+3(l-1)+k,\mu} .
\end{equation}
with
\[ U \sigma_{S}^{\mu} \otimes \sigma_{E}^{\nu} U^{\dagger} = u_{\alpha, \beta}^{\mu,\nu} \sigma_{S}^{\alpha}\otimes \sigma_{E}^{\beta}. \]

\subsection{Pairwise entangled initial state}

\begin{figure}[!htb]
		\includegraphics[scale=0.5]{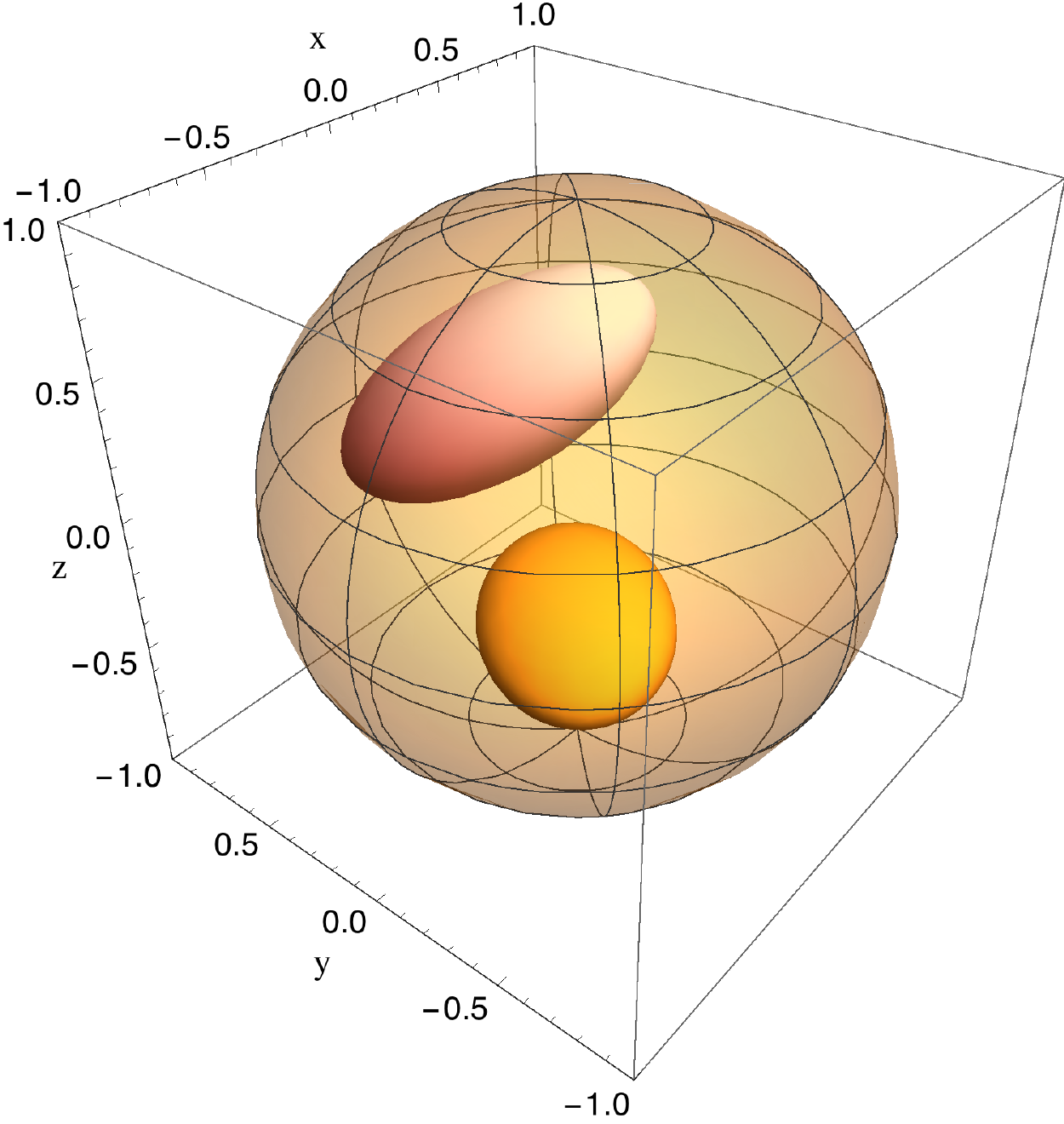}
	\caption{The reduced steering set of $S$ obtained from that of $S\!E$ generated from a tripartite $R\!S\!E$ state with $R\!S$ and $R\!E$ entanglements as given by the concurrence values $\mathcal{C}_{RS}=0.3677\, ; \mathcal{C}_{RE}=0.1102 \, ;  \mathcal{C}_{SE}=0$ is the ellipsoid (yellow) near the south pole of the Bloch sphere. This is the set on which the NCP dynamical map induced by unitary evolution of $S\!E$ with $V_{SE}=\exp[2i(\sigma_{2}^{S} \otimes\sigma_{2}^{E})]$ acts.  Here $\sigma_i^J$ are the Pauli spin operators acting on the $J^{\rm th}$ qubit. The map transforms the initial steering set to the larger (pink) ellipsoid.}
	\label{fig1}
\end{figure}

We consider a particular case wherein $\rho_{\!R\!S\!E}$ has $R\!S$ and $R\!E$ entanglement. To generate such a state, we start from a three qubit pure product state 
\[ \tau_{\!R\!S\!E} = |abc\rangle \langle abc|, \]
$a,b,c=0,1$ and apply two unitary transformations as 
\[ \rho_{\!R\!S\!E}=U_{RE}[U_{RS}\tau_{RSE}U_{RS}^{\dag}]U_{RE}^{\dag}. \]
The first operation produces $R\!S$ entanglement and the second one produces $R\!E$ entanglement. However the two operations together can entangle $S\!E$ also and so the state is passed through a controlled entanglement breaking channel to remove this entanglement. The state is brought to the canonical form in which $\rho_{R} = \openone_{R}/2$ by  the SLOCC operator $(2\rho_{R})^{-1/2} \otimes \openone_{SE}$. Finally a post-selection on to states with the desired types of entanglement is also done. The state $\rho_{\!R\!S\!E}$ could still have residual nonClassical correlations (as quantified by, for instance, the quantum Discord) between $S$ and $E$ but no entanglement. Using this procedure we generate a $R\!S\!E$ state with pairwise entanglement given by the concurrence values $\mathcal{C}_{RS}=0.3677\, ; \mathcal{C}_{RE}=0.1102 \, ;  \mathcal{C}_{SE}=0$. The $\Theta$-matrix corresponding to the numerical density matrix $\rho_{\!R\!S\!E}$ is generated and the first four elements of the vector $\vec{e}^{X} = \Theta X$ gives the reduced steering set of $S$. This reduced steering set is shown in Fig.~\ref{fig1}. Four linearly independent initial states belonging to the reduced steering set of $S$ are generated by choosing the four linearly independent values $(1,0,0)$, $(-1,0,0)$, $(0,1,0)$ and $(0,0,1)$ for $\vec{X}$ noting that the transformation that gives the corresponding state in the reduced steering set of $S$ is linear. The linear independence of the four states obtained in the reduced steering set is verified numerically to discount certain pathological scenarios. 

We now fix the $S\!E$ dynamics to be one generated by a Hamiltonian proportional to $\sigma_2^S \otimes \sigma_2^E$ and apply the unitary $\openone_{R} \otimes V_{SE}$, where $V_{SE}=\exp[2i(\sigma_{2}^{S}\otimes\sigma_{2}^{E})]$ on $\rho_{\!R\!S\!E}$ we get the transformed state $\widetilde{\rho}_{\!R\!S\!E}$ from which again the reduced steering states of $S$ can be obtained using the same procedure as above.The transformed set of states corresponding to the action of a unitary on the steering states of $S\!E$ is also shown in Fig.~\ref{fig1}. 

The four transformed states corresponding to the initial linearly independent set can easily be obtained and these initial and final states allow us to find the 16 elements of the matrix $A$ in $(\widetilde{\rho}_{S}^{X})_{ij} = A_{ij;i'j'} (\rho_{S}^{X})_{i'j'}$ by straightforward inspection.  Rearranging this matrix as $B_{ii';j'j} = A_{ij;i'j'}$ we obtain the $B$-matrix form~\cite{ecgs61b} of the dynamical map corresponding to the open evolution of $S$. The reduced dynamics in $B$-matrix form has eigenvalues $(2.3838, 0.2288, -0.5704, -0.0422)$ with the negative eigenvalues indicating that the dynamical map is NCP.). This is also seen from Fig.~\ref{fig1} in that the map transforms the reduced steering set of $S$ to a larger set and is not contractive in nature.

From this example we see that starting from a single state of $R\!S\!E$ we obtain a compact reduced steering set for $S$ which gets transformed to another such set that lies well within the state space of $S$ as a result of the possibly NCP reduced dynamics generated by an arbitrarily chosen $S\!E$ coupling. We have also outlined how such a map may be obtained through a quantum process tomography experiment that traces the evolution of sufficient number of linearly independent initial states of $S$. 

\subsection{Minimally parametrized initial $R\!S\!E$ state}

\begin{figure}[!htb]
		\includegraphics[scale=0.6]{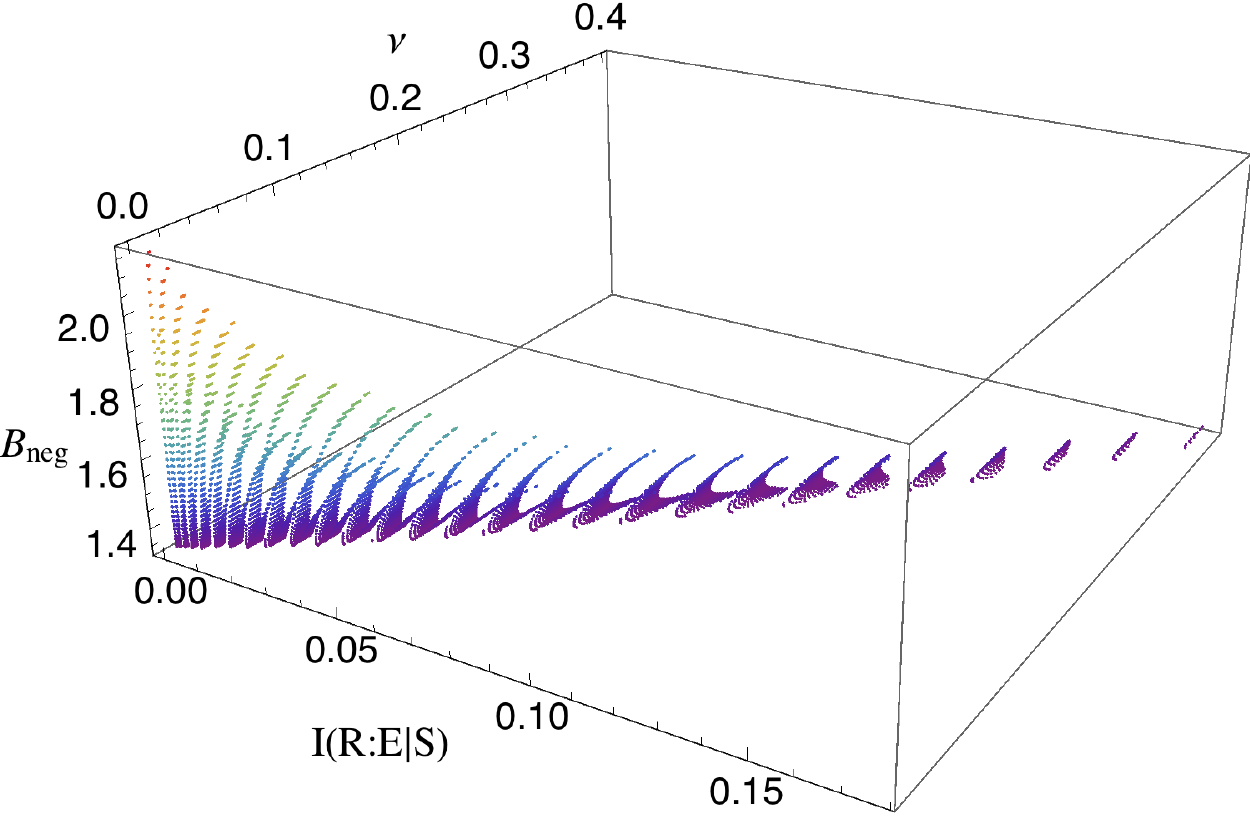}\\
	\caption{Scatter plot of $I(R:E|S)$, the DPI violation $\nu$ and the negativity of the dynamical matrix $B_{\rm neg}$ for the minimally parameterized tripartite states $\rho_{\!R\!S\!E}(P,Q,T)$ corresponding to all allowed values of $P$, $Q$ and $T$ that keep the density matrix positive. In the plot, the values of $P$, $Q$ and $T$ range from $-1$ to $1$ in steps of $0.025$. For each such $\rho_{R\!S\!E}$ generated, the DPI violation $\nu$ is maximized over the parameter $\omega$ of the unitary $S\!E$ coupling $V_{S\!E}$ and the corresponding $B_{\rm neg}$ is computed.}
	\label{fig2}
\end{figure}

A unitary transformation of the form $V_{SE}=\exp[i\omega(\sigma_{2}\otimes\sigma_{2})]$ acting on the $S\!E$ states of the type given in Eq.~(\ref{eq:rhoSE2}) induces the following transformation on the Bloch vector components of $S$:
\begin{eqnarray*}
	e_{1}^{X} & \rightarrow & e_{1}^{X} \cos(2 \omega) - e_{14}^{X} \sin (2 \omega) \\
	e_{2}^{X}  & \rightarrow &  e_{2}^{X} \\
	e_{3}^{X}  & \rightarrow & e_{3}^{X} \cos (2 \omega) + e_{8}^{X} \sin (2 \omega).
\end{eqnarray*} 
Since the transformation depends only on $e_{8}^{X}$ and $e_{14}^{X}$ apart from the Bloch vector components of $S$, we consider next a minimally parametrized $R\!S\!E$ state that produce a nontrivial dynamical map by setting 
	\begin{eqnarray*}
		e_{1} = e_{3} & = &  P, \\
		e_{5} & = & Q, \\
		e_{8} = e_{14} & = &  E, \\
		T_{2,i}= T_{5,i}=T_{8,i}=T_{14,i} & = &  T
	\end{eqnarray*}
and all other parameters in Eq.~(\ref{eq:rhoRSE2}) equal to zero. Having $E=PQ$ ensures $S\!E$ separability in $\rho_{\!R\!S\!E}$, while $T_{2,i}$ and $T_{5,i}$ ensure that the initial state has nonClassical correlations in the $R\!S$ and $R\!E$ subsystems. For this three parameter family of states $\rho_{\!R\!S\!E}(P,Q,T)$ with  
\[ e_{8}^{X} = e_{14}^{X} = PQ + T(X_{1} + X_{2} + X_{3}),\] we can systematically study the  interdependence of the Markovianity of the initial tripartite state as quantified by $I(R:E|S)$, the degree of violation of the quantum DPI and the NCP nature of the reduced dynamics. The degree of violation of DPI is quantified by the difference between the mutual information between $R$ and $S$ before and after the joint $S\!E$ evolution under $V_{SE}$ as  $\nu = \max (0, I(R:S') - I(R:S))$ and the NCP nature of the map is quantified by the {\em negativity} of the $B$ matrix defined as 
\[ B_{\rm neg} =  \sum_{j} |\lambda_{j}|-2,\]
where $\lambda_{j}$ are the eigenvalues of the $B$ matrix and its trace is 2. In Fig.~\ref{fig2} a scatter plot of these three quantities for $\rho_{\!R\!S\!E}(P,Q,T)$ for all allowed values of $B$, $C$ and $T$ is shown. For each such initial state the parameter $\omega$ in $V_{SE}$ is chosen so as to maximise $\nu$. We see that the violation of the quantum DPI is more when there is more delocalised information between $R$ and $E$ in the initial system indicating that this shared information can flow to $S$ during the joint evolution of $S$ and $E$. In all the cases considered, the open evolution of $S$ is NCP but the negativity of the $B$ matrix does not seem to have a definite relationship to $I(R:E|S)$ and $\nu$ at least for the set of unitary transformations $V_{SE}$ considered. This is primarily because we are considering only a restricted set of unitary transformations on the $S\!E$ system. The maximisation of the DPI in this case is only over $\omega$ and not over all possible unitary transformations of the $S\!E$ subsystem. The fixed choice for the type of unitary transformation limits the scope of redistribution of the initial delocalised information during the course of the dynamics. The particular dynamics that brings maximum possible shared information between $R$ and $E$ into $S$ is not the one that has the largest negativity of the reduced dynamics of $S$ but it still is NCP in nature. 

\subsection{Random initial $R\!S\!E$ states}

\begin{figure}[!htb]
		\includegraphics[scale=0.65]{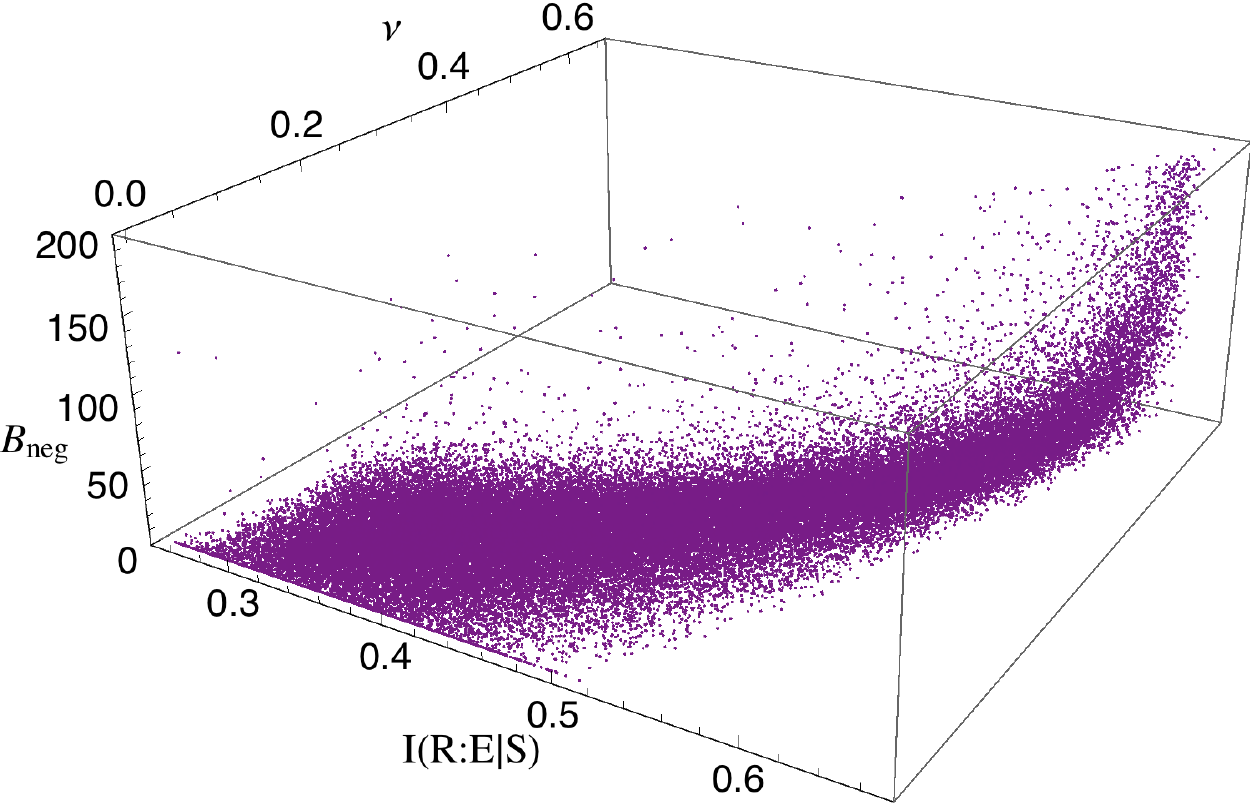}
	\caption{Scatter plot of $I(R:E|S)$, the DPI violation $\nu$ and the negativity of the dynamical matrix $B_{\rm neg}$ for around 70,000 randomly generated states $\rho_{\!R\!S\!E}$ with no initial $S\!E$ entanglement. The DPI violation is maximised over 3000 random unitaries each}
	\label{fig3}
\end{figure}

In our third numerical example, 70,000 initial $\rho_{R\!S\!E}$ states are generated with no $SE$ entanglement using the same procedure as with the first example. After applying a randomly generated $S\!E$ unitary acting as $\openone_{R} \otimes V_{SE}$ on $\rho_{\!R\!S\!E}$ we compute the DPI violation using the initial state and the transformed state $\widetilde{\rho}_{\!R\!S\!E}$. We maximize the quantum DPI violation over 3000 instances of randomly generated two qubit unitary matrices for each randomly generated initial state. The $B$-matrix form of the map corresponding to the unitary that gives the maximal violation of the quantum DPI is computed and its negativity is calculated.  In Fig.~\ref{fig3} a scatter plot of $I(R:E|S)$ of the initial state against the maximal violation, $\nu$ of the quantum DPI and the negativity of the corresponding $B$-matrix for the 70,000 trials is shown. Here we see that $\nu$ is higher for higher values of $I(R:E|S)$ in the intitial state and $B_{\rm neg}$ is large when both $I(R:E|S)$ and $\nu$ have high values.

From Fig.~\ref{fig3} we see that as the DPI violation increases and as the initial mutual information between $R$ and $E$ increases, the negativity of the $B$-matrix also rises on an average. Generation of the random initial states fails around thirty percent of the time to produce a state with the desired kinds of nonClassical correlations (only between $R\!S$ and $R\!E$). The successful trials also fails less than one percent of the time because it leads to a reduced steering set in the Bloch sphere of states of $S$ that is one or two dimensional. In these cases, the strategy of reconstructing the $B$-matrix by taking four linearly independent projectors on $R$ to produce four linearly independent states in the reduced steering set of $S$ fails. These failed trials are characterized by non-numerical or non-Hermitian $B$ matrices based on which such trials are weeded out. Reduced steering sets that are very small also leads to numerical errors that typically lead to very large (order of $10^{2}$)  values for $B_{\rm neg}$. These are also removed by limiting the range over which $B_{\rm neg}$ is plotted.

\begin{figure}[!htb]
\resizebox{7.5 cm}{14 cm}{\includegraphics{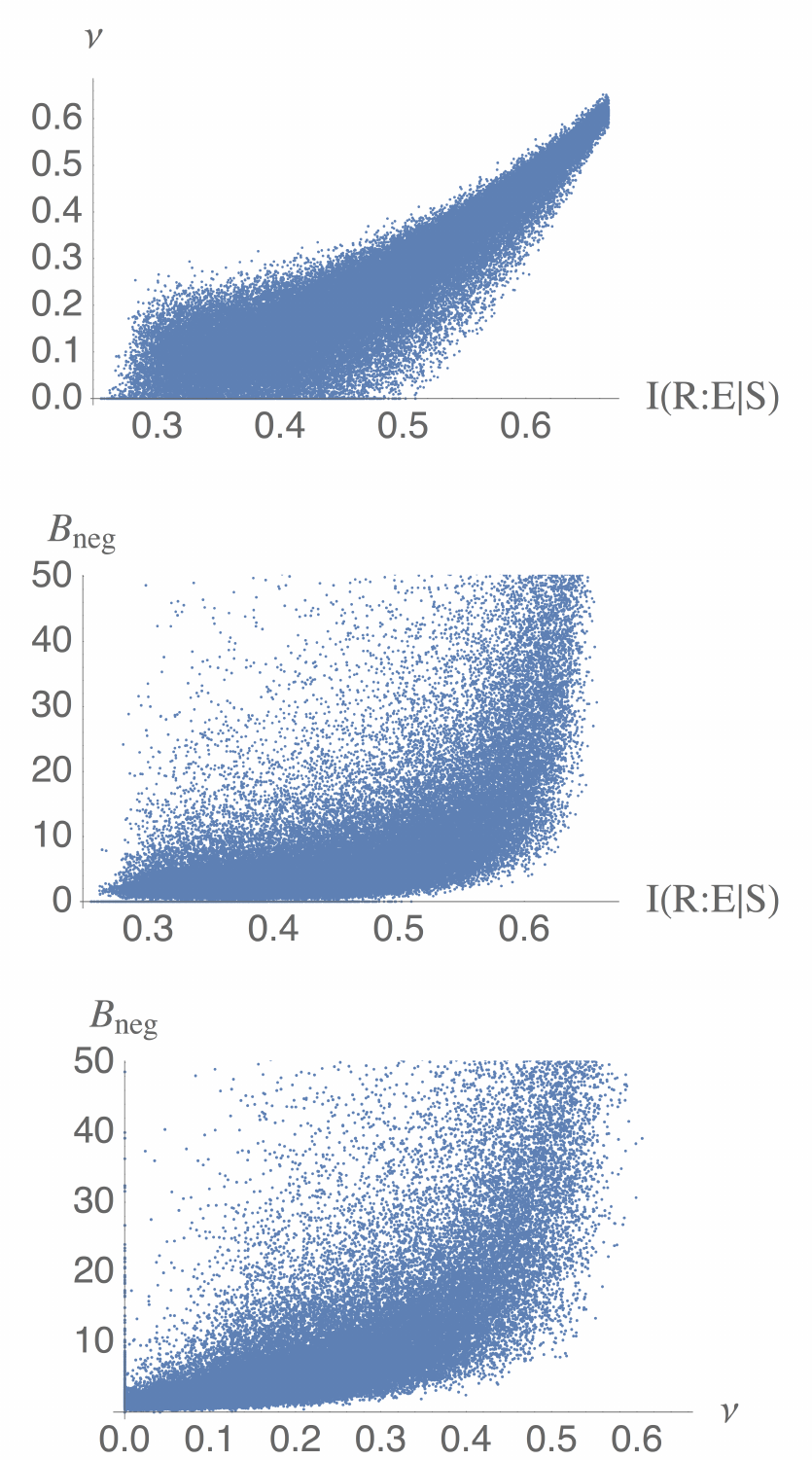}}
\caption{Projections of the scatter plot given in Fig~(\ref{fig3}) for around 70,000 randomly generated states showing the inter-relationships between $I(R:E|S)$, $\nu$ and $B_{\rm neg}$. \label{figAB2}}
\end{figure}

Sections of the scatter plot in Fig.~(\ref{fig3}) are given in Fig.~(\ref{figAB2}). Unlike in the second example we see clear connections between the three quantities $I(R:E|S)$, $B_{\rm neg}$ and $\nu$. As before the violation of the quantum DPI increases with increasing $I(R:E|S)$ of the initial state. The corresponding $B_{\rm neg}$ is also seen to increase. From the last panel in Fig.~(\ref{figAB2}) we see that larger violations of the quantum DPI typically tend to generate NCP dynamical maps with greater negativity. The numerical data suggests lower and upper bounds on each of the quantities as functions of the others. These bounds remain to be explored analytically.

\section{Conclusion \label{sec:summary}}

The reference system in the form of either a preparation device or the rest of the universe is a ubiquitous and unavoidable element in the analysis of any quantum process tomography experiment. Under only very specific conditions can we completely ignore the effects of the delocalized quantum information that can exist across the system its environment and the reference. For discussing what could show up as a result of a quantum process tomography experiment we limited $R$ to be the preparation device. Ideal preparation devices and/or very large environments on which both $R$ and $S$ has negligible effect can guarantee an initial product state of $S\!E$ leading to CP maps defined on the entire state space of $S$. The ideal preparation device will allow one to initialize the system of interest in any desired initial state and in particular into a complete linearly independent set of initial states on which the effect of the environment can be studied. After interaction with the large environment, state tomography on the final states corresponding to the initial set would reveal a CP dynamical map. 

However if the preparation device is non-ideal in that it cannot remove all initial correlations between $S$ and $E$ then it can initialize $S$ only into a subset of available states. Sufficient number of linearly independent initial states may still be accessible to experimentally reconstruct a dynamical map, which then is likely to be NCP but still positivity preserving on the subset of states that can be prepared by $R$. Only in those subset of cases where the initial imperfect preparation in no way affects $E$ do we get a CP dynamical map as in~\cite{Buscemi:2014dc} because $R\!S\!E$ forms a short Markov chain. With recent advances in the control and manipulation of single quantum systems like for instance, a single atom in a crystal or an ion among few others, the immediate environment is also mostly quantum in nature consisting of a few other atoms, photons or ions. In this case the non-ideal classical preparation device is more likely than not affecting $E$ as well as $S$ during the initial preparation. In this case also sufficient number of linearly independent initial states of the $S$ would be accessible to enable quantum process tomography. But as shown in our main result the dynamical map reconstructed from the observations is going to be NCP but with a very well defined interpretation and domain of action. 

In the present Paper, $R$, $S$ and $E$ are limited to be finite dimensional quantum systems. Extension to cases where the subsystems; in particular $R$ and $E$ are infinite dimensional remains to be done. However there are no assumptions made in proving the main result that suggest that it may not be applicable in the case where either $E$ or $R$ or both are quantum systems with a continuum of levels. Prior to extending our formal proof to such cases, we first have to extend the idea of steering ellipsoids to those cases. Since doing so, in itself, would form a substantial body of work, we leave it as work to be done in the future. As far as finite dimensional systems are concerned, our main result is a statement about the nature of the initial state of the system and the environment augmented by the presence of a passive reference system. This means that the open quantum dynamics obtained for each one of the large number of models of $S\!E$ interaction available in the literature~\cite{breuer02} can be re-computed by adding the reference and considering interesting initial $R\!S\!E$ states that are plausible within the assumptions of the model to obtain the conditions under which NCP dynamics will be observed. There are no changes to the $S\!E$ coupling and dynamics in the models since $R$ remains passive.

It may be noted that a careful analysis of reported quantum tomography experiments like for instance~\cite{howard_quantum_2006} reveals that often NCP maps are actually suggested by data but the conceptual difficulties that existed previously regarding interpreting such dynamics meant that the observed map was approximated with a suitable CP dynamical map. The results presented in this Paper clears up all these conceptual and technical difficulties regarding NCP dynamics.

\section*{Acknowledgements}

This work is supported in part by a grant from SERB, DST, Government of India No. EMR/2016/007221 and by the Ministry of Human Resources Development, Government of India through the FAST program.

\bibliography{DPIandNCP}

\begin{thebibliography}{35}%
\makeatletter
\providecommand \@ifxundefined [1]{%
 \@ifx{#1\undefined}
}%
\providecommand \@ifnum [1]{%
 \ifnum #1\expandafter \@firstoftwo
 \else \expandafter \@secondoftwo
 \fi
}%
\providecommand \@ifx [1]{%
 \ifx #1\expandafter \@firstoftwo
 \else \expandafter \@secondoftwo
 \fi
}%
\providecommand \natexlab [1]{#1}%
\providecommand \enquote  [1]{``#1''}%
\providecommand \bibnamefont  [1]{#1}%
\providecommand \bibfnamefont [1]{#1}%
\providecommand \citenamefont [1]{#1}%
\providecommand \href@noop [0]{\@secondoftwo}%
\providecommand \href [0]{\begingroup \@sanitize@url \@href}%
\providecommand \@href[1]{\@@startlink{#1}\@@href}%
\providecommand \@@href[1]{\endgroup#1\@@endlink}%
\providecommand \@sanitize@url [0]{\catcode `\\12\catcode `\$12\catcode
  `\&12\catcode `\#12\catcode `\^12\catcode `\_12\catcode `\%12\relax}%
\providecommand \@@startlink[1]{}%
\providecommand \@@endlink[0]{}%
\providecommand \url  [0]{\begingroup\@sanitize@url \@url }%
\providecommand \@url [1]{\endgroup\@href {#1}{\urlprefix }}%
\providecommand \urlprefix  [0]{URL }%
\providecommand \Eprint [0]{\href }%
\providecommand \doibase [0]{http://dx.doi.org/}%
\providecommand \selectlanguage [0]{\@gobble}%
\providecommand \bibinfo  [0]{\@secondoftwo}%
\providecommand \bibfield  [0]{\@secondoftwo}%
\providecommand \translation [1]{[#1]}%
\providecommand \BibitemOpen [0]{}%
\providecommand \bibitemStop [0]{}%
\providecommand \bibitemNoStop [0]{.\EOS\space}%
\providecommand \EOS [0]{\spacefactor3000\relax}%
\providecommand \BibitemShut  [1]{\csname bibitem#1\endcsname}%
\let\auto@bib@innerbib\@empty
\bibitem [{\citenamefont {Breuer}\ and\ \citenamefont
  {Petruccione}(2002)}]{breuer02}%
  \BibitemOpen
  \bibfield  {author} {\bibinfo {author} {\bibfnamefont {H.~P.}\ \bibnamefont
  {Breuer}}\ and\ \bibinfo {author} {\bibfnamefont {F.}~\bibnamefont
  {Petruccione}},\ }\href@noop {} {\emph {\bibinfo {title} {The theory of open
  quantum systems}}}\ (\bibinfo  {publisher} {Oxford university press},\
  \bibinfo {address} {New York},\ \bibinfo {year} {2002})\BibitemShut {NoStop}%
\bibitem [{\citenamefont {Lloyd}(2013)}]{Lloyd:2013wi}%
  \BibitemOpen
  \bibfield  {author} {\bibinfo {author} {\bibfnamefont {S.}~\bibnamefont
  {Lloyd}},\ }\href@noop {} {\enquote {\bibinfo {title} {{The universe as
  quantum computer}},}\ } (\bibinfo {year} {2013}),\ \Eprint
  {http://arxiv.org/abs/arXiv:1312.4455} {arXiv:1312.4455} \BibitemShut
  {NoStop}%
\bibitem [{\citenamefont {Ingarden}\ \emph {et~al.}()\citenamefont {Ingarden},
  \citenamefont {Kossakowski},\ and\ \citenamefont {Ohya}}]{Ingarden:vw}%
  \BibitemOpen
  \bibfield  {author} {\bibinfo {author} {\bibfnamefont {R.~S.}\ \bibnamefont
  {Ingarden}}, \bibinfo {author} {\bibfnamefont {A.}~\bibnamefont
  {Kossakowski}}, \ and\ \bibinfo {author} {\bibfnamefont {M.}~\bibnamefont
  {Ohya}},\ }\href@noop {} {\emph {\bibinfo {title} {{Information Dynamics and
  Open Systems}}}}\ (\bibinfo  {publisher} {Springer Netherlands})\BibitemShut
  {NoStop}%
\bibitem [{\citenamefont {Sudarshan}\ \emph {et~al.}(1961)\citenamefont
  {Sudarshan}, \citenamefont {Mathews},\ and\ \citenamefont {Rau}}]{ecgs61b}%
  \BibitemOpen
  \bibfield  {author} {\bibinfo {author} {\bibfnamefont {E.~C.~G.}\
  \bibnamefont {Sudarshan}}, \bibinfo {author} {\bibfnamefont {P.~M.}\
  \bibnamefont {Mathews}}, \ and\ \bibinfo {author} {\bibfnamefont
  {J.}~\bibnamefont {Rau}},\ }\href@noop {} {\bibfield  {journal} {\bibinfo
  {journal} {Phys. Rev.}\ }\textbf {\bibinfo {volume} {121}},\ \bibinfo {pages}
  {920} (\bibinfo {year} {1961})}\BibitemShut {NoStop}%
\bibitem [{\citenamefont {Choi}(1972)}]{choi72}%
  \BibitemOpen
  \bibfield  {author} {\bibinfo {author} {\bibfnamefont {M.~D.}\ \bibnamefont
  {Choi}},\ }\href@noop {} {\bibfield  {journal} {\bibinfo  {journal} {Can. J.
  Math.}\ }\textbf {\bibinfo {volume} {24}},\ \bibinfo {pages} {520} (\bibinfo
  {year} {1972})}\BibitemShut {NoStop}%
\bibitem [{\citenamefont {Kraus}(1971)}]{kraus71a}%
  \BibitemOpen
  \bibfield  {author} {\bibinfo {author} {\bibfnamefont {K.}~\bibnamefont
  {Kraus}},\ }\href@noop {} {\bibfield  {journal} {\bibinfo  {journal} {Ann.
  Phys.}\ }\textbf {\bibinfo {volume} {64}},\ \bibinfo {pages} {311} (\bibinfo
  {year} {1971})}\BibitemShut {NoStop}%
\bibitem [{\citenamefont {Kraus}(1983)}]{kraus83}%
  \BibitemOpen
  \bibfield  {author} {\bibinfo {author} {\bibfnamefont {K.}~\bibnamefont
  {Kraus}},\ }\href@noop {} {\emph {\bibinfo {title} {States, Effects and
  Operations: Fundamental notions of Quantum Theory}}},\ edited by\ \bibinfo
  {editor} {\bibfnamefont {A.}~\bibnamefont {Bohm}}, \bibinfo {editor}
  {\bibfnamefont {J.~D.}\ \bibnamefont {Dollard}}, \ and\ \bibinfo {editor}
  {\bibfnamefont {W.~H.}\ \bibnamefont {Wooters}},\ \bibinfo {series} {Lecture
  notes in Physics}, Vol.\ \bibinfo {volume} {190}\ (\bibinfo  {publisher}
  {Spring-Verlag},\ \bibinfo {address} {New York},\ \bibinfo {year}
  {1983})\BibitemShut {NoStop}%
\bibitem [{\citenamefont {Nielsen}\ and\ \citenamefont
  {Chuang}(2000)}]{nielsen00}%
  \BibitemOpen
  \bibfield  {author} {\bibinfo {author} {\bibfnamefont {M.~A.}\ \bibnamefont
  {Nielsen}}\ and\ \bibinfo {author} {\bibfnamefont {I.~L.}\ \bibnamefont
  {Chuang}},\ }\href@noop {} {\emph {\bibinfo {title} {Quantum computation and
  quantum information}}}\ (\bibinfo  {publisher} {Cambridge university press},\
  \bibinfo {address} {Cambridge, U. K.},\ \bibinfo {year} {2000})\BibitemShut
  {NoStop}%
\bibitem [{\citenamefont {Stinespring}(1955)}]{stinespring55}%
  \BibitemOpen
  \bibfield  {author} {\bibinfo {author} {\bibfnamefont {W.~F.}\ \bibnamefont
  {Stinespring}},\ }\href@noop {} {\bibfield  {journal} {\bibinfo  {journal}
  {Proc. Amer. Math. Soc.}\ }\textbf {\bibinfo {volume} {6}},\ \bibinfo {pages}
  {211} (\bibinfo {year} {1955})}\BibitemShut {NoStop}%
\bibitem [{\citenamefont {Shumacher}(1996)}]{schumacher96a}%
  \BibitemOpen
  \bibfield  {author} {\bibinfo {author} {\bibfnamefont {B.}~\bibnamefont
  {Shumacher}},\ }\href@noop {} {\bibfield  {journal} {\bibinfo  {journal}
  {Phys. Rev. A}\ }\textbf {\bibinfo {volume} {54}},\ \bibinfo {pages} {2614}
  (\bibinfo {year} {1996})}\BibitemShut {NoStop}%
\bibitem [{\citenamefont {Jordan}\ \emph {et~al.}(2004)\citenamefont {Jordan},
  \citenamefont {Shaji},\ and\ \citenamefont {Sudarshan}}]{jordan04}%
  \BibitemOpen
  \bibfield  {author} {\bibinfo {author} {\bibfnamefont {T.~F.}\ \bibnamefont
  {Jordan}}, \bibinfo {author} {\bibfnamefont {A.}~\bibnamefont {Shaji}}, \
  and\ \bibinfo {author} {\bibfnamefont {E.~C.~G.}\ \bibnamefont {Sudarshan}},\
  }\href@noop {} {\bibfield  {journal} {\bibinfo  {journal} {Phys. Rev A.}\
  }\textbf {\bibinfo {volume} {70}},\ \bibinfo {pages} {052110} (\bibinfo
  {year} {2004})}\BibitemShut {NoStop}%
\bibitem [{\citenamefont {Jordan}(2004)}]{jordan04b}%
  \BibitemOpen
  \bibfield  {author} {\bibinfo {author} {\bibfnamefont {T.~F.}\ \bibnamefont
  {Jordan}},\ }\href@noop {} {\bibfield  {journal} {\bibinfo  {journal} {Phys.
  Rev. A}\ }\textbf {\bibinfo {volume} {71}},\ \bibinfo {pages} {034101}
  (\bibinfo {year} {2004})}\BibitemShut {NoStop}%
\bibitem [{\citenamefont {Pechukas}(1994)}]{pechukas94}%
  \BibitemOpen
  \bibfield  {author} {\bibinfo {author} {\bibfnamefont {P.}~\bibnamefont
  {Pechukas}},\ }\href@noop {} {\bibfield  {journal} {\bibinfo  {journal}
  {Phys. Rev. Lett.}\ }\textbf {\bibinfo {volume} {73}},\ \bibinfo {pages}
  {1060} (\bibinfo {year} {1994})}\BibitemShut {NoStop}%
\bibitem [{\citenamefont {Pechukas}(1995)}]{pechukas95}%
  \BibitemOpen
  \bibfield  {author} {\bibinfo {author} {\bibfnamefont {P.}~\bibnamefont
  {Pechukas}},\ }\href@noop {} {\bibfield  {journal} {\bibinfo  {journal}
  {Phys. Rev. Lett.}\ }\textbf {\bibinfo {volume} {75}},\ \bibinfo {pages}
  {3021} (\bibinfo {year} {1995})}\BibitemShut {NoStop}%
\bibitem [{\citenamefont {Shaji}\ and\ \citenamefont
  {Sudarshan}(2005)}]{shaji05a}%
  \BibitemOpen
  \bibfield  {author} {\bibinfo {author} {\bibfnamefont {A.}~\bibnamefont
  {Shaji}}\ and\ \bibinfo {author} {\bibfnamefont {E.~C.~G.}\ \bibnamefont
  {Sudarshan}},\ }\href@noop {} {\bibfield  {journal} {\bibinfo  {journal}
  {Phys. Lett. A.}\ }\textbf {\bibinfo {volume} {341}},\ \bibinfo {pages} {48}
  (\bibinfo {year} {2005})}\BibitemShut {NoStop}%
\bibitem [{\citenamefont {Brodutch}\ \emph {et~al.}(2013)\citenamefont
  {Brodutch}, \citenamefont {Datta}, \citenamefont {Modi}, \citenamefont
  {Rivas},\ and\ \citenamefont {Rodr{\'\i}guez-Rosario}}]{Brodutch:2013bx}%
  \BibitemOpen
  \bibfield  {author} {\bibinfo {author} {\bibfnamefont {A.}~\bibnamefont
  {Brodutch}}, \bibinfo {author} {\bibfnamefont {A.}~\bibnamefont {Datta}},
  \bibinfo {author} {\bibfnamefont {K.}~\bibnamefont {Modi}}, \bibinfo {author}
  {\bibfnamefont {{\'A}.}~\bibnamefont {Rivas}}, \ and\ \bibinfo {author}
  {\bibfnamefont {C.~A.}\ \bibnamefont {Rodr{\'\i}guez-Rosario}},\ }\href@noop
  {} {\bibfield  {journal} {\bibinfo  {journal} {Phys.~Rev.~A}\ }\textbf
  {\bibinfo {volume} {87}},\ \bibinfo {pages} {042301} (\bibinfo {year}
  {2013})}\BibitemShut {NoStop}%
\bibitem [{\citenamefont {Dominy}\ \emph {et~al.}(2015)\citenamefont {Dominy},
  \citenamefont {Shabani},\ and\ \citenamefont {Lidar}}]{Dominy:2015fw}%
  \BibitemOpen
  \bibfield  {author} {\bibinfo {author} {\bibfnamefont {J.~M.}\ \bibnamefont
  {Dominy}}, \bibinfo {author} {\bibfnamefont {A.}~\bibnamefont {Shabani}}, \
  and\ \bibinfo {author} {\bibfnamefont {D.~A.}\ \bibnamefont {Lidar}},\
  }\href@noop {} {\bibfield  {journal} {\bibinfo  {journal} {Quantum
  Information Processing}\ }\textbf {\bibinfo {volume} {15}},\ \bibinfo {pages}
  {465} (\bibinfo {year} {2015})}\BibitemShut {NoStop}%
\bibitem [{\citenamefont {Rodriguez}\ \emph {et~al.}(2008)\citenamefont
  {Rodriguez}, \citenamefont {Modi}, \citenamefont {Kuah}, \citenamefont
  {Shaji},\ and\ \citenamefont {Sudarshan}}]{rodriguez_completely_2008}%
  \BibitemOpen
  \bibfield  {author} {\bibinfo {author} {\bibfnamefont {C.~A.}\ \bibnamefont
  {Rodriguez}}, \bibinfo {author} {\bibfnamefont {K.}~\bibnamefont {Modi}},
  \bibinfo {author} {\bibfnamefont {A.-m.}\ \bibnamefont {Kuah}}, \bibinfo
  {author} {\bibfnamefont {A.}~\bibnamefont {Shaji}}, \ and\ \bibinfo {author}
  {\bibfnamefont {E.~C.~G.}\ \bibnamefont {Sudarshan}},\ }\href {\doibase
  10.1088/1751-8113/41/20/205301} {\bibfield  {journal} {\bibinfo  {journal}
  {J. Phys. A: Math. Theor.}\ }\textbf {\bibinfo {volume} {41}},\ \bibinfo
  {pages} {205301} (\bibinfo {year} {2008})}\BibitemShut {NoStop}%
\bibitem [{\citenamefont {Shabani}\ and\ \citenamefont
  {Lidar}(2009)}]{Shabani:2009dy}%
  \BibitemOpen
  \bibfield  {author} {\bibinfo {author} {\bibfnamefont {A.}~\bibnamefont
  {Shabani}}\ and\ \bibinfo {author} {\bibfnamefont {D.~A.}\ \bibnamefont
  {Lidar}},\ }\href@noop {} {\bibfield  {journal} {\bibinfo  {journal} {Phys.
  Rev. Lett.}\ }\textbf {\bibinfo {volume} {102}},\ \bibinfo {pages} {100402}
  (\bibinfo {year} {2009})}\BibitemShut {NoStop}%
\bibitem [{\citenamefont {Shabani}\ and\ \citenamefont
  {Lidar}(2016)}]{Shabani:2016if}%
  \BibitemOpen
  \bibfield  {author} {\bibinfo {author} {\bibfnamefont {A.}~\bibnamefont
  {Shabani}}\ and\ \bibinfo {author} {\bibfnamefont {D.~A.}\ \bibnamefont
  {Lidar}},\ }\href@noop {} {\bibfield  {journal} {\bibinfo  {journal} {Phys.
  Rev. Lett.}\ }\textbf {\bibinfo {volume} {116}},\ \bibinfo {pages} {049901}
  (\bibinfo {year} {2016})}\BibitemShut {NoStop}%
\bibitem [{\citenamefont {Buscemi}(2014)}]{Buscemi:2014dc}%
  \BibitemOpen
  \bibfield  {author} {\bibinfo {author} {\bibfnamefont {F.}~\bibnamefont
  {Buscemi}},\ }\href@noop {} {\bibfield  {journal} {\bibinfo  {journal}
  {Phys.~Rev.~Lett.}\ }\textbf {\bibinfo {volume} {113}},\ \bibinfo {pages}
  {140502} (\bibinfo {year} {2014})}\BibitemShut {NoStop}%
\bibitem [{\citenamefont {Breuer}\ \emph {et~al.}(2009)\citenamefont {Breuer},
  \citenamefont {Laine},\ and\ \citenamefont {Piilo}}]{Breuer:2009cu}%
  \BibitemOpen
  \bibfield  {author} {\bibinfo {author} {\bibfnamefont {H.-P.}\ \bibnamefont
  {Breuer}}, \bibinfo {author} {\bibfnamefont {E.-M.}\ \bibnamefont {Laine}}, \
  and\ \bibinfo {author} {\bibfnamefont {J.}~\bibnamefont {Piilo}},\
  }\href@noop {} {\bibfield  {journal} {\bibinfo  {journal} {Phys. Rev. Lett.}\
  }\textbf {\bibinfo {volume} {103}},\ \bibinfo {pages} {210401} (\bibinfo
  {year} {2009})}\BibitemShut {NoStop}%
\bibitem [{\citenamefont {D{\textquoteright}Arrigo}\ \emph
  {et~al.}(2014)\citenamefont {D{\textquoteright}Arrigo}, \citenamefont
  {Lo~Franco}, \citenamefont {Benenti}, \citenamefont {Paladino},\ and\
  \citenamefont {Falci}}]{DArrigo:2014jd}%
  \BibitemOpen
  \bibfield  {author} {\bibinfo {author} {\bibfnamefont {A.}~\bibnamefont
  {D{\textquoteright}Arrigo}}, \bibinfo {author} {\bibfnamefont
  {R.}~\bibnamefont {Lo~Franco}}, \bibinfo {author} {\bibfnamefont
  {G.}~\bibnamefont {Benenti}}, \bibinfo {author} {\bibfnamefont
  {E.}~\bibnamefont {Paladino}}, \ and\ \bibinfo {author} {\bibfnamefont
  {G.}~\bibnamefont {Falci}},\ }\href@noop {} {\bibfield  {journal} {\bibinfo
  {journal} {Annals of Physics}\ }\textbf {\bibinfo {volume} {350}},\ \bibinfo
  {pages} {211} (\bibinfo {year} {2014})}\BibitemShut {NoStop}%
\bibitem [{\citenamefont {Jennings}\ and\ \citenamefont
  {Rudolph}(2010)}]{Jennings:2010kr}%
  \BibitemOpen
  \bibfield  {author} {\bibinfo {author} {\bibfnamefont {D.}~\bibnamefont
  {Jennings}}\ and\ \bibinfo {author} {\bibfnamefont {T.}~\bibnamefont
  {Rudolph}},\ }\href@noop {} {\bibfield  {journal} {\bibinfo  {journal}
  {Phys.~Rev.~E}\ }\textbf {\bibinfo {volume} {81}},\ \bibinfo {pages} {061130}
  (\bibinfo {year} {2010})}\BibitemShut {NoStop}%
\bibitem [{\citenamefont {Partovi}(2008)}]{Partovi:2008if}%
  \BibitemOpen
  \bibfield  {author} {\bibinfo {author} {\bibfnamefont {M.~H.}\ \bibnamefont
  {Partovi}},\ }\href@noop {} {\bibfield  {journal} {\bibinfo  {journal}
  {Phys.~Rev.~E}\ }\textbf {\bibinfo {volume} {77}},\ \bibinfo {pages} {021110}
  (\bibinfo {year} {2008})}\BibitemShut {NoStop}%
\bibitem [{\citenamefont {Xu}\ \emph {et~al.}(2013)\citenamefont {Xu},
  \citenamefont {Sun}, \citenamefont {Li}, \citenamefont {Xu}, \citenamefont
  {Guo}, \citenamefont {Andersson}, \citenamefont {Lo~Franco},\ and\
  \citenamefont {Compagno}}]{Xu:2013ia}%
  \BibitemOpen
  \bibfield  {author} {\bibinfo {author} {\bibfnamefont {J.-S.}\ \bibnamefont
  {Xu}}, \bibinfo {author} {\bibfnamefont {K.}~\bibnamefont {Sun}}, \bibinfo
  {author} {\bibfnamefont {C.-F.}\ \bibnamefont {Li}}, \bibinfo {author}
  {\bibfnamefont {X.-Y.}\ \bibnamefont {Xu}}, \bibinfo {author} {\bibfnamefont
  {G.-C.}\ \bibnamefont {Guo}}, \bibinfo {author} {\bibfnamefont
  {E.}~\bibnamefont {Andersson}}, \bibinfo {author} {\bibfnamefont
  {R.}~\bibnamefont {Lo~Franco}}, \ and\ \bibinfo {author} {\bibfnamefont
  {G.}~\bibnamefont {Compagno}},\ }\href@noop {} {\bibfield  {journal}
  {\bibinfo  {journal} {Nature Comm.}\ }\textbf {\bibinfo {volume} {4}}
  (\bibinfo {year} {2013})}\BibitemShut {NoStop}%
\bibitem [{\citenamefont {Rivas}\ \emph {et~al.}(2014)\citenamefont {Rivas},
  \citenamefont {Huelga},\ and\ \citenamefont {Plenio}}]{Rivas:2014bl}%
  \BibitemOpen
  \bibfield  {author} {\bibinfo {author} {\bibfnamefont {{\'A}.}~\bibnamefont
  {Rivas}}, \bibinfo {author} {\bibfnamefont {S.~F.}\ \bibnamefont {Huelga}}, \
  and\ \bibinfo {author} {\bibfnamefont {M.~B.}\ \bibnamefont {Plenio}},\
  }\href@noop {} {\bibfield  {journal} {\bibinfo  {journal} {Rep.~Prog.~Phys.}\
  }\textbf {\bibinfo {volume} {77}},\ \bibinfo {pages} {094001} (\bibinfo
  {year} {2014})}\BibitemShut {NoStop}%
\bibitem [{\citenamefont {Wiseman}\ \emph {et~al.}(2007)\citenamefont
  {Wiseman}, \citenamefont {Jones},\ and\ \citenamefont
  {Doherty}}]{Wiseman:2007ca}%
  \BibitemOpen
  \bibfield  {author} {\bibinfo {author} {\bibfnamefont {H.~M.}\ \bibnamefont
  {Wiseman}}, \bibinfo {author} {\bibfnamefont {S.~J.}\ \bibnamefont {Jones}},
  \ and\ \bibinfo {author} {\bibfnamefont {A.~C.}\ \bibnamefont {Doherty}},\
  }\href@noop {} {\bibfield  {journal} {\bibinfo  {journal} {Phys. Rev. Lett.}\
  }\textbf {\bibinfo {volume} {98}},\ \bibinfo {pages} {140402} (\bibinfo
  {year} {2007})}\BibitemShut {NoStop}%
\bibitem [{\citenamefont {Cavalcanti}\ and\ \citenamefont
  {Skrzypczyk}(2017)}]{Cavalcanti:2017ba}%
  \BibitemOpen
  \bibfield  {author} {\bibinfo {author} {\bibfnamefont {D.}~\bibnamefont
  {Cavalcanti}}\ and\ \bibinfo {author} {\bibfnamefont {P.}~\bibnamefont
  {Skrzypczyk}},\ }\href@noop {} {\bibfield  {journal} {\bibinfo  {journal}
  {Rep.~Prog.~Phys.}\ }\textbf {\bibinfo {volume} {80}},\ \bibinfo {pages}
  {024001} (\bibinfo {year} {2017})}\BibitemShut {NoStop}%
\bibitem [{\citenamefont {Modi}(2012)}]{Modi:2012iv}%
  \BibitemOpen
  \bibfield  {author} {\bibinfo {author} {\bibfnamefont {K.}~\bibnamefont
  {Modi}},\ }\href@noop {} {\bibfield  {journal} {\bibinfo  {journal}
  {Scientific Reports}\ }\textbf {\bibinfo {volume} {2}} (\bibinfo {year}
  {2012})}\BibitemShut {NoStop}%
\bibitem [{\citenamefont {Jevtic}\ \emph {et~al.}(2014)\citenamefont {Jevtic},
  \citenamefont {Pusey}, \citenamefont {Jennings},\ and\ \citenamefont
  {Rudolph}}]{Jevtic:2014fm}%
  \BibitemOpen
  \bibfield  {author} {\bibinfo {author} {\bibfnamefont {S.}~\bibnamefont
  {Jevtic}}, \bibinfo {author} {\bibfnamefont {M.}~\bibnamefont {Pusey}},
  \bibinfo {author} {\bibfnamefont {D.}~\bibnamefont {Jennings}}, \ and\
  \bibinfo {author} {\bibfnamefont {T.}~\bibnamefont {Rudolph}},\ }\href@noop
  {} {\bibfield  {journal} {\bibinfo  {journal} {Phys.~Rev.~Lett.}\ }\textbf
  {\bibinfo {volume} {113}},\ \bibinfo {pages} {020402} (\bibinfo {year}
  {2014})}\BibitemShut {NoStop}%
\bibitem [{\citenamefont {Chuang}\ and\ \citenamefont
  {Nielsen}(2009)}]{Chuang:2009ie}%
  \BibitemOpen
  \bibfield  {author} {\bibinfo {author} {\bibfnamefont {I.~L.}\ \bibnamefont
  {Chuang}}\ and\ \bibinfo {author} {\bibfnamefont {M.~A.}\ \bibnamefont
  {Nielsen}},\ }\href@noop {} {\bibfield  {journal} {\bibinfo  {journal}
  {J.~Mod.~Opt.}\ } (\bibinfo {year} {2009})}\BibitemShut {NoStop}%
\bibitem [{\citenamefont {Yuen-Zhou}(2014)}]{YuenZhou:2014js}%
  \BibitemOpen
  \bibfield  {author} {\bibinfo {author} {\bibfnamefont {J.}~\bibnamefont
  {Yuen-Zhou}},\ }in\ \href@noop {} {\emph {\bibinfo {booktitle} {Ultrafast
  Spectroscopy}}}\ (\bibinfo  {publisher} {IOP Publishing},\ \bibinfo {year}
  {2014})\BibitemShut {NoStop}%
\bibitem [{\citenamefont {Zurek}(2003)}]{Zurek:2003fm}%
  \BibitemOpen
  \bibfield  {author} {\bibinfo {author} {\bibfnamefont {W.~H.}\ \bibnamefont
  {Zurek}},\ }\href@noop {} {\bibfield  {journal} {\bibinfo  {journal} {Reviews
  of Modern Physics}\ }\textbf {\bibinfo {volume} {75}},\ \bibinfo {pages}
  {715} (\bibinfo {year} {2003})}\BibitemShut {NoStop}%
\bibitem [{\citenamefont {Howard}\ \emph {et~al.}(2006)\citenamefont {Howard},
  \citenamefont {Twamley}, \citenamefont {Wittmann}, \citenamefont {Gaebel},
  \citenamefont {Jelezko},\ and\ \citenamefont
  {Wrachtrup}}]{howard_quantum_2006}%
  \BibitemOpen
  \bibfield  {author} {\bibinfo {author} {\bibfnamefont {M.}~\bibnamefont
  {Howard}}, \bibinfo {author} {\bibfnamefont {J.}~\bibnamefont {Twamley}},
  \bibinfo {author} {\bibfnamefont {C.}~\bibnamefont {Wittmann}}, \bibinfo
  {author} {\bibfnamefont {T.}~\bibnamefont {Gaebel}}, \bibinfo {author}
  {\bibfnamefont {F.}~\bibnamefont {Jelezko}}, \ and\ \bibinfo {author}
  {\bibfnamefont {J.}~\bibnamefont {Wrachtrup}},\ }\href {\doibase
  10.1088/1367-2630/8/3/033} {\bibfield  {journal} {\bibinfo  {journal} {New
  Journal of Physics}\ }\textbf {\bibinfo {volume} {8}},\ \bibinfo {pages} {33}
  (\bibinfo {year} {2006})}\BibitemShut {NoStop}%
\end{thebibliography}%


\end{document}